\documentstyle[11pt]{article}
\input epsf

\newlength{\awidth}
\newlength{\aheight}
\newlength{\uswidth}
\newlength{\usheight}
\def\preprint#1{\gdef\@preprint{#1}}
\def\bce{\begin{center}}
\def\ece{\end{center}}
\def\be{\begin{equation}}
\def\ee{\end{equation}}
\def\bea{\begin{eqnarray}}
\def\eea{\end{eqnarray}}

\newcounter{fignr}

\begin{document}
\baselineskip=.285in

\catcode`\@=11
\def\maketitle{\par
 \begingroup
 \def\thefootnote{\fnsymbol{footnote}}
 \def\@makefnmark{\hbox
 to 0pt{$^{\@thefnmark}$\hss}}
 \if@twocolumn
 \twocolumn[\@maketitle]
 \else \newpage
 \global\@topnum\z@ \@maketitle \fi\thispagestyle{empty}\@thanks
 \endgroup
 \setcounter{footnote}{0}
 \let\maketitle\relax
 \let\@maketitle\relax
 \gdef\@thanks{}\gdef\@author{}\gdef\@title{}\let\thanks\relax}
\def\@maketitle{\newpage
 \null
 \hbox to\textwidth{\hfil\hbox{\begin{tabular}{r}\@preprint\end{tabular}}}
 \vskip 2em \begin{center}
 {\Large\bf \@title \par} \vskip 1.5em {\normalsize \lineskip .5em
\begin{tabular}[t]{c}\@author
 \end{tabular}\par}
 \end{center}
 \par
 \vskip 1.5em}
\def\preprint#1{\gdef\@preprint{#1}}
\def\abstract{\if@twocolumn
\section*{Abstract}
\else \normalsize
\begin{center}
{\large\bf Abstract\vspace{-.5em}\vspace{0pt}}
\end{center}
\quotation
\fi}
\def\endabstract{\if@twocolumn\else\endquotation\fi}
\catcode`\@=12

\preprint{}
\title{\Large\bf Extracting Gravitational Energy From The Homogeneous Isotropic Universe
\protect\\[1mm]\  }
\author{\normalsize Eue Jin Jeong\\[1mm]
{\normalsize\it Department of Physics, Natural Science Research Institute, Yonsei
University, Seoul, Korea}}

\maketitle

\renewcommand{\theequation}{\thesection.\arabic{equation}}
\def\gatij{\gamma^{ij}}
\def\gabij{\gamma_{ij}}
\def\ophi{\phi^{a}}
\def\hphi{\overline{\phi}^{a}}

\begin{center}
{\large\bf Abstract}\\[3mm]
\end{center}
\indent\indent
The kinetic energy of a local system of objects placed in a curved spacetime is 
gained by the subsequent acceleration of the object following the more
contracted region of spacetime. Normally this happens near massive gravitating 
stars. However, the gravitational dipole moment has been shown to be capable of 
self creating asymmetrically distorted spacetime in its vicinity, thereby, 
capable of being accelerated indefinitely following the successive self 
created loophole of the spacetime. Localization of this kinetic energy 
may be possible by designing a system that uses the artificially 
created gravitational dipole moments to rotate the main axis. 
A mechanical constraint is derived for the extraction of unlimited 
gravitational energy from such system.

\vspace{1cm}\noindent
Keyword(s): gravitational energy; curved spacetime; general relativity; 
dipole gravity; rotating hemisphere

\noindent
PACS number(s): 03.30.+p, 04.30.+x

\baselineskip=15pt

\pagenumbering{arabic}
\thispagestyle{plain}
\setcounter{section}{1}

\vspace{1cm}\indent

It has been discussed in the previous paper that a longitudinal axially 
asymmetric rotating object violates NewtonÕs first and third laws of 
motion. In this system, the angular degree of freedom of motion is coupled 
to that of the linear one since the angular momentum 
produces the anomalous displacement of the center of mass along 
the direction of the rotation axis. This is a totally unexpected 
phenomenon from the view point of Newtonian mechanics. 
The translational gauge symmetry in the 
quantized version of general relativity is spontaneously 
broken for the rotating hemispherical system since the usual 
constant phase factor that shifts the coordinate system without 
affecting the energy content of the system is no longer the constant 
in such cases. 
It has to carry the information about the kinetic energy of the rotor 
which depends on the time derivative of the angular orientation $d{\theta}/dt$ 
as well as the exact geometrical shape of the rotor to know the 
exact effective center of mass of 
the object at any given moment of time. The result of this breaking 
of symmetry is the non conservation of energy and linear momentum which are 
the direct consequences of Newton's first and third laws of motion. 
As is well known, all of the six degrees of freedom 
of motion(3 translational and 3 rotational) of an object must be independent 
from each other in Newtonian mechanics. And its validity has never 
been challenged in general relativity either. 
Any one of the degrees of freedom of motion should not affect any 
other degrees of freedom of motion, which seems not the case for the 
mentioned system.  
It seems that the spacetime has lesser degree of freedom in such systems 
since the linear degree of freedom of motion is 
somehow intermingled with the time dimension through the angular frequency of 
rotation of the rotor. This new physical phenomenon has the potential to affect 
QM and EM and especially the 
Thermodynamics since the kinetic theory of gas has never scrutinized the 
longitudinal axially asymmetric type of gas like methane in the light of 
this new principle. The measured values of $C_{v}$ 
and $\gamma$ have revealed that there is no consistent law governing them\cite{B1} 
for the methane derivative gases like $CH_{3}Cl$, $CH_{2}Cl_{2}$, $CHCl_{3}$, 
$CCl_{4}$ which are of the tetrahedral shape that can have the 
longitudinal axially asymmetric rotational modes. Thus,the 
$C_{v}$ and $\gamma$ of the above gases may need to be recalculated 
in the light of the present theory and compared with the previous experimental data.

In general, linear displacement of an object occurs typically in 
mechanics in three cases. First, the coordinate translation, 
second the constant rectilinear motion with respect to the stationary coordinate 
system and third due to the accelerated motion of an object. 
The coordinate translation does not really describe the motion of an 
object. ItÕs only a shift of the reference point. 
The second case is the typical rectilinear motion described by NewtonÕs 
first law of motion. 
The third case occurs when the object is subjected to a time rated change of 
the external action where the acting force is given by 
$d(mv)/dt$. In the first two cases, external energy is not required 
to change the orientation of the object. In the third case, the 
displacement is a function of the external force and the mass of 
the object. In case the object is subjected to a uniform external force, 
the displacement is the second order function of time. And in most 
of the other similar cases, NewtonÕs second and third law of motion 
can effectively deal with the mechanics of the system where the energy 
and momentum of the total system are conserved.

There is yet another kind of displacement of an object which depends 
on the internal energy of the system. The idealized, perfectly rigid 
rotating hemispherical rotor which is an example case of 
longitudinal axially asymmetric rotor has been considered extensively in 
the previous paper which generates the shift of the center of mass 
perpendicular to the plane of rotation. The length element defined as 
the "anomalous center of mass shift" arising from the energy dependent 
center of mass has the following peculiar characteristics. 1. It 
is related to an internal energy of the system. 2. The larger the 
shift of the center of mass, the greater the stored energy. 3. It 
can be returned to zero upon releasing the related internal energy. 
As can be expected, this anomalous center of mass shift is 
independent of the coordinate system. Since the breaking of 
NewtonÕs first law motion indicates the possibility of 
spontaneous linear movement of the object, it is expected 
that this local system will gain energy as it moves along. 

The possibility of extracting vacuum energy by using the attractive 
Casimir force between metal layers has been proposed by Forward\cite{B2} in 
1984. More recently, Cole and Puthoff\cite{B3} have again raised the possibility 
of energy extraction from the vacuum. While the Casimir force is the 
manifestation of the electromagnetic zero point fields, the vacuum 
as contractible spacetime seems to have other interesting properties in 
energetics purely from the relativistic point of view.

The dipole gravitational moment represents a new physical entity 
that defies Newton's first and third laws of motion in which the local 
energy is not conserved, which is similar to the case of a 
local system of objects placed in a non flat spacetime in general 
relativity\cite{B4}\cite{B5}\cite{B6}\cite{B7}. 
It is physically conceivable that the gravitational dipole moment given by \\ \\
\[\int T^{oo} x'_{j} \,d^3 x' = M\delta
\bar{r}_{c}\] \\ \\ where $T^{oo}$ is the total mass energy density of 
the rotating source and $\delta
\bar{r}_{c}$ the anomalous center of mass shift vector, would interact with 
the gravitational potential field pervading the universe and be directionally 
accelerated in the same way that an electric dipole moment would be propelled 
toward certain direction depending on its original position and orientation 
in a sparsely populated electrostatic charges in a non conducting 
spherical shell. Calculation of the actual amount of force using 
the distance and matter distribution in the real universe by 
integration is hampered by the fact that not only the universe 
is not in the three dimensional manifold but also that there is no 
confirmed data for the exact amount of total mass in the universe. 
Instead, one may be able to calculate the total mass of the universe using 
the measured linear force in the three dimensional approximation. 

As discussed in the previous paper, if we choose the proposition that the 
center of the centripetal force 
tries to stay at the rest state center of mass due to the inertial 
resistance while the centrifugal 
force exerts force centering around the shifted center of mass, the net 
result is that there 
remains non zero vertical component of force in the hemispherical 
system with respect to the rest of the universe. 
Then the linear force can be calculated to be, by using the triangular law, \\ \\
\[F_{linear} = F_{centrifugal}(\frac{\delta r_{c}}{\sqrt{2/3} R})\] \\ \\
where $\delta r_{c}$ is the anomalous shift of the center of mass 
and $\sqrt{2/3} R$ the effective radius where the total mass is imagined 
to be concentrated while giving the same inertia as that of the hemisphere 
for $R\omega << c$. And subsequently, \\ \\
 \[F_{linear} = \frac{\pi m\omega^4 R^3}
{48\sqrt{\frac{2}{3}} c^2}\] \\ \\
for $R\omega << c$, where R is the radius of the hemispherical shell, m the 
mass and ${\omega}$ the angular frequency of the rotor respectively. 

Once the rotational motion of an idealized perfectly rigid object 
which is axisymmetric yet longitudinal axially asymmetric is 
proven to be capable of creating a locally asymmetric spacetime 
distortion and the corresponding force is given by the above 
expression where the force depends on the fourth power of the 
angular frequency ${\omega}$, the next step of using this force to rotate 
the wheel(on the axis of which a generator may be attached to 
produce electricity) is straight forward.

It must be noted at this point that there exists one important, well known, 
mechanical 
constraint in this process. The dipole rotor has to overcome the 
resisting force acting against changing the direction of its own 
angular momentum in the process of performing the work to rotate the 
main axis. Since 
overcoming this resisting force would require energy to be drawn 
from the system, the energy generated by the dipole rotor must be 
greater than the energy required to change its own angular momentum 
plus all other forms of energy loss for positive energy production. 

To determine the mechanical constraint for positive energy production 
following the above discussion, consider a device which has the shape 
of a large scale classic wind speedometer with four arms of equal 
length attached perpendicular to the main axis horizontally stretched 
90 degree to each other. The axis of four rotating hemispheres are 
attached at the end of each arms perpendicular to both the main axis 
and the arms respectively. 

Consider an infinitesimal distance 
$dS=rd{\theta}$ traveled by the dipole 
rotor attached at the end of the arms of the device due to the 
force exerted on itself. Assuming that all other moving components in 
the device are massless except one dipole rotor which is activated 
and massive, the amount of work exercised on the wheel during the 
infinitesimal travel is given by
\\ \\ \[\overline{F}\cdot d \overline{S}=
\frac{\pi m\omega^4 R^3}{48\sqrt{\frac{2}{3}} c^2}r
d{\theta}\]\\ \\ where R and ${\omega}$ are the radius and the angular frequency
of the hemispherical rotor respectively and r the length and $d\theta$ 
the infinitesimal angular rotation of the arms. The energy spent to change 
the angular orientation of the dipole moment is given approximately for 
$R\omega << c$ by \\ \\
\[{\tau}d{\theta}=\frac{dL}{dt}d{\theta}=|\overline{\Omega}
\times\overline{L}|d{\theta}={\Omega}Ld{\theta}={\Omega}\frac{2}{3}mR^2
{\omega}d{\theta}\]\\ \\ where 
$\Omega$ is the angular frequency of the main axis in the system. 
Even with the assumption that all the frictional energy loss can 
be eliminated completely, this is the fundamental low limit of 
the energy loss required to make up by the force exerted on the 
dipole. 
The output energy must be greater than this fundamental energy 
loss, so that the condition \\ \\ \[\frac{\pi m\omega^4 R^3}
{48\sqrt{\frac{2}{3}} c^2}rd{\theta}\geq\Omega\frac{2}{3}m
R^2{\omega}d{\theta}\]\\ \\
must be satisfied, which gives 
\\ \\ \[\frac{\omega^3 R r}
{8.3 c^2}\geq \Omega\]\\ \\
for the idealized perfectly rigid rotating hemispherical shell for $R\omega/c 
<< 1$. This clearly demonstrates that this localized ideal system 
is capable of producing positive energy. The ${\Omega}$ sets the maximum 
available angular frequency for given R, r and ${\omega}$. In the normal 
stabilized energy production mode, the ${\Omega}$ would slow down and maintain 
the smaller value than the one given by the above condition depending 
on how much energy is drawn from the system. The total amount of 
energy produced depends on the sixth power of the angular frequency 
${\omega}$ and on the second power of r and R respectively. 

It is 
possible that the energy created here now may be lost somewhere some other time 
in the universe in such a way that the total mass energy of the entire 
universe remains always constant, although it's an uncomfortable 
conjecture that may never be proved. Still, on the surface, the energy 
can be obtained only when the homogeneous isotropic universe is assumed to 
be filled with matter, exerting long range gravitational 
interactions, not in the universe which is totally void.

\def\hebibliography#1{\begin{center}\subsection*{References}
\end{center}\list
  {[\arabic{enumi}]}{\settowidth\labelwidth{[#1]}
\leftmargin\labelwidth	  \advance\leftmargin\labelsep
    \usecounter{enumi}}
    \def\newblock{\hskip .11em plus .33em minus .07em}
    \sloppy\clubpenalty4000\widowpenalty4000
    \sfcode`\.=1000\relax}

\let\endhebibliography=\endlist

\begin{hebibliography}{100}

\bibitem{B1} J. H. Jeans, D. Sc., LL. D. and F. R. S., The Dynamical Theory of 
Gases (Cambridge University Press, London, 1925) P. 192
\bibitem{B2} R. L. Forward, Phys. Rev. B {\bf 30}, 1700 (1984)
\bibitem{B3} Daniel C. Cole and H. E. Puthoff, Phys. Rev. E {\bf 48}, 1562 (1993)
\bibitem{B4} M. Alcubierre, Class. and Quantum Grav. {\bf 11}, L73 (1994)
\bibitem{B5} A. Einstein, Sitzungsber. Preuss. Akad. Wiss. Phys. Math. {\bf K1}, 
688 (1916); 154 (1918).
\bibitem{B6} E. Noether, Nachr. Ges. Wiss. Goettingen {\bf 2}, 235(1918) ; J. G. 
Fletcher, Rev. Mod. Phys. {\bf 32}, 65(1960) ; R. Wald, J. Math. Phys. 
{\bf 31}, 2378 (1990) ; Dongsu Bak, D. Cangemi, and R. Jackiw, Phys. Rev. 
D. {\bf 49}, 5173 (1994)
\bibitem{B7} C. W. Misner, K. S. Thorne and J. A. Wheeler, Gravitation 
(Freeman, San Francisco, 1973)

\end{hebibliography}
\end{document}